\documentclass[aps,prb,twocolumn,superscriptaddress,showpacs]{revtex4} 
\usepackage[dvips]{graphicx}
\usepackage{bm}% bold math
 
\usepackage{latexsym} 
\usepackage{amsmath}
\usepackage{amssymb}

\usepackage{bm}

\newcommand{\bmr}{{\bm r}}

\newcommand{\bmq}{{\bm q}}

\newcommand{\bmk}{{\bm k}}

\begin{document}

\title{
Spin transport in half-metallic ferromagnets
}

\author{Y. Ohnuma}
%\email{ohnuma.yuichi@jaea.go.jp}
\affiliation{%
Advanced Science Research Center, Japan Atomic Energy Agency, Tokai 319-1195, Japan
}% 
\affiliation{%
ERATO, Japan Science and Technology Agency, Sendai, Miyagi 980-8577, Japan
}% 

\author{M. Matsuo}
\affiliation{%
Advanced Science Research Center, Japan Atomic Energy Agency, Tokai 319-1195, Japan
}%
\affiliation{%
ERATO, Japan Science and Technology Agency, Sendai, Miyagi 980-8577, Japan
}%

\author{S. Maekawa}
\affiliation{%
Advanced Science Research Center, Japan Atomic Energy Agency, Tokai 319-1195, Japan
}%
\affiliation{%
ERATO, Japan Science and Technology Agency, Sendai, Miyagi 980-8577, Japan
}%

%\pacs{} 
\date{\today}

\begin{abstract} 
We theoretically investigate spin transport in half-metallic ferromagnets at finite temperatures. The side-jump and skew-scattering contributions to spin Hall conductivity are derived using the Kubo formula. The electron-magnon interaction causes a finite density of states in the energy gap of the minority-spin band and induces spin Hall conductivity. We show that spin Hall conductivity is proportional to $T^{3/2}$, with $T$ being temperature and is sensitive to $T$. We propose that spin Hall conductivity may be a tool to study the minority-spin state. 
\end{abstract} 
%\keywords{} 
\maketitle %\maketitle must follow title, authors, abstract and \pacs
%%%%%%%%%%%%%%%%%%%%%%%%%%%%%%%%%%%%%%%%%%%%%%%%%%%%%%%%%%%%%%%%%%%%%%%
%%%%%%%%%%%%%%%%%%%%%%%%%%%%%%%%%%%%%%%%%%%%%%%%%%%%%%%%%%%%%%%%%%%%%%%
%===
\section{Introduction} 
%===
Since the discovery of giant magnetoresistance in 1988, much attention has been paid to the spin transport in the field of spintronics~\cite{Zutic04, Maekawa-text}. In particular, a half-metallic ferromagnet, where the conduction electrons are completely spin polarized, is regarded as a  candidate material for a spin injector in magnetic memories~\cite{Katsnelson08, Takanashi10, Felser07}. Although a 100\% spin-polarized ferromagnetic metal has yet to be discovered, high spin polarization has been observed in some Heusler alloys at room temperature~\cite{Felser07, Sakuraba06, Chioncel08, Ikhtiar16}-this makes these alloys an attractive option for further applications.

The spin polarization of a half-metallic ferromagnet has been investigated theoretically. De Groot {\it et al~}~\cite{Groot83} showed that the first principle calculation yields that 100\% spin polarization is realized with a minority-spin band gap. 
However, some theoretical investigations proposed that the spin polarization should be suppressed by the electron-electron correlation~\cite{Hertz72, Hertz73, Edwards73, Edwards83, Irkhin06}. Unlike the spin polarization, however, spin transport in half-metallic ferromagnets has yet to be studied. 

In this paper, we theoretically investigate spin transport with an electron-electron correlation in half-metallic ferromagnets. Specific attention is paid to the extrinsic spin Hall effect, wherein a spin current flows perpendicular to an applied electric field in the presence of spin-orbit scattering caused by impurities~\cite{Sinova16, Hoffmann13}. Extrinsic spin Hall conductivity is calculated with the Kubo formula and it is shown that at finite temperatures, minority-spin electrons contribute to spin Hall conductivity owing to thermally excited magnons. The efficiency of the pure spin injections into a half-metallic ferromagnet is also discussed. The temperature dependence of spin Hall conductivity and spin polarization is compared, and it is shown that while spin Hall conductivity drastically increases with temperature the spin polarization remains mostly constant. This suggests that the observation of spin Hall conductivity may become a method for studying the minority-spin state in a half-metallic ferromagnet. 

The outline of the paper is as follows.
In Sec.~\ref{Sec:RevHMF}, a brief review of the effect of the electron-electron correlation on the density of states in a half-metallic ferromagnet is given. 
In Sec.~\ref{Sec:NLSV_HMF}, spin injection into a half-metallic ferromagnet in lateral spin valve structures is discussed. 
In Sec.~\ref{Sec:SHE}, extrinsic spin Hall conductivity is derived. The temperature dependence of spin Hall conductivity is discussed by comparing with that of spin polarization. It is shown that spin Hall conductivity is more sensitive to temperature than spin polarization.
In Sec.~\ref{Sec:Conclusion}, we summarize our results. 
%===
\section{Density of states in half-metallic ferromagnets \label{Sec:RevHMF}}
%===
In this section, a brief review of the effect of the electron-electron correlation on the density of states in a half-metallic ferromagnet is given.
Following Refs.~[\onlinecite{Hertz72, Hertz73, Edwards73, Irkhin06}], we start with the Hubbard model to describe the electron-electron correlation in a half-metallic ferromagnet,
%%%
\begin{eqnarray}
  H_{\rm Hub} 
	&=& 
  \sum_{\bmk}\sum_{\alpha}\varepsilon_{\bmk\alpha} c^{\dag}_{\bmk\alpha}c_{\bmk\alpha} 
  + U\sum_{i}c^{\dag}_{i\uparrow}c_{i\uparrow}c^{\dag}_{i\downarrow}c_{i\downarrow},
\label{Eq:Model-Hub}
\end{eqnarray}
%%%		
where $c^{\dag}_{\bmk\alpha}$ and $c_{\bmk\alpha}$ are the electron creation and annihilation operators with spin polarization $\alpha=\uparrow,\downarrow$, $\varepsilon_{\bmk\alpha}$ is the kinetic energy of electrons, and $U$ is the on-site Coulomb repulsion.

\begin{figure}[h] 
		\begin{center}
		\scalebox{0.3}[0.3]{\includegraphics{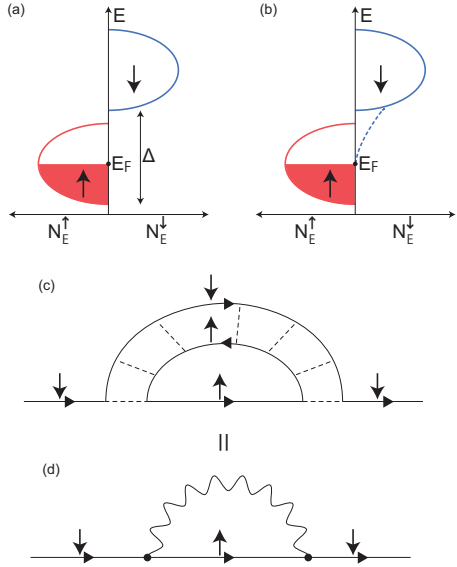}}
		\end{center}
		\caption{(Color online)
			(a) Schematic view of the spin-dependent density of states of a half-metallic ferromagnet in the Hartree approximation. Here, $\Delta$ is the exchange band splitting. (b) As shown by the dotted curve, the gap closes owing to electron-electron correlation~\cite{Edwards73}. (c) Feynman diagram of the self-energy of the minority-spin electrons in terms of electron-electron scattering~\cite{Hertz72}. (d) Feynman diagram of the self-energy of the minority-spin electrons in terms of electron-magnon scattering~\cite{Hertz72}. The solid and wavy lines in (c) and (d) represent the electron and magnon propagators, respectively, while the dashed lines describe the Coulomb repulsion.  
		}
		\label{fig1_Ohnuma}
	\end{figure}

The Hartree approximation in the Hubbard model (\ref{Eq:Model-Hub}) yields that $\varepsilon_{\bmk\alpha}$ for $\alpha=\uparrow,\downarrow$ is given by
$\varepsilon_{\bmk\uparrow}=\varepsilon_{\bmk}$
and
$\varepsilon_{\bmk\downarrow}=\varepsilon_{\bmk} + \Delta$, where 
$\Delta=Un_{\uparrow}$ is the exchange band splitting, with $n_{\uparrow}$ being the electron number density with $\alpha=\uparrow$. 
Figure~\ref{fig1_Ohnuma}(a) illustrates the spin-dependent density of states of a half-metallic ferromagnet in the Hartree approximation, and it can be seen from this figure that exchange band splitting $\Delta$ opens a gap in the minority-spin band. 
%%%%%%%%%%%%%%%%%%%%%%%%%%%%
%%%%%%%%%%%%%%%%%%%%%%%%%%%

Theoretical studies have shown that the gap in the minority-spin band closes owing to electron-electron correlation~\cite{Hertz72, Hertz73, Edwards73, Edwards83, Irkhin06}. 
Figure~\ref{fig1_Ohnuma}(c) shows the self-energy of the minority-spin electrons describing electron-electron correlation~\cite{Hertz72, Hertz73, Edwards73}. Absorbing the electron-hole scatterings into magnon propagators, we obtain the spin-flip process owing to magnon excitation, as shown in Fig.~\ref{fig1_Ohnuma}(d)~\cite{Hertz72, Hertz73}. 
The self-energy of the minority-spin electrons, ${\rm Im}\Sigma^{R, U}_{\bmk\downarrow}$ as shown in Fig.~\ref{fig1_Ohnuma}(b), can then be given by
%%%
\begin{eqnarray}
	{\rm Im}\Sigma^{R, U}_{\bmk\downarrow}
	&=& -\frac{\pi U\Delta}{N} \sum_{\bmq}(1-f_{\varepsilon_{\bmk+\bmq\uparrow}} + n_{\omega_{\bmq}}) \nonumber \\
	&\times&\delta(\varepsilon_{\rm F} - \varepsilon_{\bmk+\bmq\uparrow} - \hbar\omega_{\bmq}),
\label{Eq:SelfE-Hub}
\end{eqnarray}
%%%		
where $\omega_{\bmq}=D\bmq^2$ is the magnon dispersion, with $D$ being the stiffness constant, while $f_{\varepsilon}$ and $n_{\omega}$ describe the Fermi-Dirac and the Bose-Einstein distribution functions, respectively. 
The density of states of the minority-spin electrons $N^{\downarrow}_{\varepsilon}$ at zero temperature, shown in Fig.~\ref{fig1_Ohnuma}(b), is given by $N^{\downarrow}_{\varepsilon}=\frac{1}{n_{\uparrow}}\sum_{\bmk'\bmq}(1-f_{\varepsilon_{\bmk'\uparrow}}+n_{\omega_{\bmq}})\delta(\varepsilon_{\bmk'}+\hbar Dq^2 - \varepsilon)$. After taking the summation over $\bmk'$ and $\bmq$, the following is obtained:
%%%
\begin{eqnarray}
	N^{\downarrow}_{\varepsilon}
	&=& N^{\uparrow}_{\varepsilon} \Big(\frac{\varepsilon - \varepsilon_{\rm F}}{\hbar Dk^2_{\rm F}}\Big)^{\frac{3}{2}}.
\label{Eq:DOS-01}
\end{eqnarray}
%%%	
Here $N^{\uparrow}_{\varepsilon}$ is the density of states of the majority-spin electrons, $\varepsilon_{\rm F}$ is the Fermi energy, and $k_{\rm F}$ is the Fermi wave number~\cite{Edwards73}. The $(\varepsilon - \varepsilon_{\rm F})^{3/2}$ law in Eq.~(\ref{Eq:DOS-01}) arises from the energy conservation associated with the majority-spin electrons and magnons~\cite{Edwards73}. Note that the density of states of the minority-spin electrons $N^{\downarrow}_{\varepsilon}$ vanishes at the Fermi level. 

At finite temperatures, the density of states of the minority-spin electrons at the Fermi level becomes finite owing to the thermally excited magnons~\cite{Edwards83, Irkhin06} and is given by
%%%
\begin{eqnarray}
	N^{\downarrow}_{\varepsilon_{\rm F}}
	&=& \frac{3}{2} N^{\uparrow}_{\varepsilon_{\rm F}} \Big(\frac{k_{\rm B} T}{\hbar Dk^2_{\rm F}}\Big)^{\frac{3}{2}}
	\gamma_0,
\label{Eq:DOS-02}
\end{eqnarray}
%%%	
where $\gamma_0\equiv\int^{\infty}_0 dx x^{1/2}(\frac{1}{e^x+1}+\frac{1}{e^x-1})=(4-\sqrt{2})\sqrt{\pi}\zeta(3/2)/4$, with $\zeta(3/2)=2.612$. The $T^{3/2}$ dependence of the density of states $N^{\downarrow}_{\varepsilon_{\rm F}}$ in Eq.~(\ref{Eq:DOS-02}) originates from the thermal excitation of the magnons~\cite{Irkhin06}. 
%===
\section{Spin injection into half-metallic ferromagnets \label{Sec:NLSV_HMF}}
%===
In this section, we consider the methods that can be used to detect the spin Hall effect in a half-metallic ferromagnet. 
To detect the spin Hall effect, a pure spin current needs to be generated~\cite{Sinova16, Hoffmann13, Saitoh06, Valenzuela06, Kimura07, Uchida08, Uchida10, Jaworski10}. There are typically three methods for generating pure spin current in metals: (a) by spin pumping~\cite{Saitoh06, Tserkovnyak05}; (b) by utilizing the spin Seebeck effect in a magnetic bilayer system~\cite{Uchida08, Uchida10, Jaworski10, Xiao10, Adachi13}; or  (c) by spin injection in a lateral spin valve structure~\cite{Maekawa-text, Valenzuela06, Kimura07, Zhang00, Takahashi08}. In spin pumping and spin Seebeck effect, pure spin current is injected into a metal using a microwave and a temperature gradient in an attached ferromagnet, while in the lateral spin valve structure, pure spin current is generated from spin accumulation in the metal. Now the metal in all three cases is replaced by a half-metallic ferromagnet. In cases (a) and (b), observed signals are determined by both spin Hall conductivity in the half-metallic ferromagnet and interaction at the interface, while in case (c), the signals are determined only by spin Hall conductivity. Therefore, for simplicity, we examine the lateral spin valve structure described by case (c) in order to detect the spin Hall effect in a half-metallic ferromagnet. 

As shown in Eq.~(\ref{Eq:DOS-01}), at zero temperature, the 100\% spin-polarized electrons ($\uparrow$ spin electron) flow so that the propagation of the pure spin current is prohibited. At finite temperatures, however, the minority-spin electrons can flow and so a pure spin current is obtained, as shown in Eq.~(\ref{Eq:DOS-02}). 
%%%%%%%%%%%%%%%%%%%%%%%%%%%%
	\begin{figure}[h] 
		\begin{center}
		\scalebox{0.5}[0.5]{\includegraphics{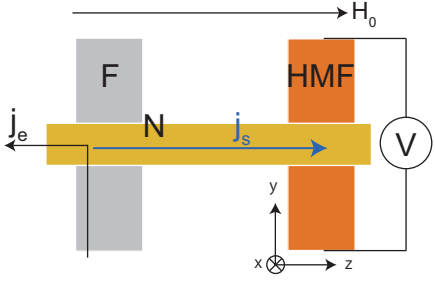}}
		\end{center}
		\caption{(Color online)
			Schematic view of a lateral spin valve structure for spin injection~\cite{Maekawa-text, Valenzuela06, Kimura07, Zhang00, Takahashi08}. A ferromagnetic wire (F) and a half-metallic ferromagnetic wire (HMF) are connected by a nonmagnetic metal (N) bridge in the $z$ direction. A pure spin current $j_s$ is generated in N wire by the charge current $j_e$ between F and N wires. The pure spin current $j_s$ flowing into N is absorbed into HMF wire in the $x$ direction. The absorbed spin current is then converted into charge current owing to its spin-orbit interaction, and the electric voltage $V$ is obtained in the $y$ direction. The magnetic field $H_0$ is applied in the $z$ direction.
		}
		\label{fig2_Ohnuma}
	\end{figure}
%%%%%%%%%%%%%%%%%%%%%%%%%%%		

Figure~\ref{fig2_Ohnuma} shows a schematic view of the lateral spin valve structure for spin injection~\cite{Maekawa-text, Valenzuela06, Kimura07, Zhang00, Takahashi08}. The pure spin current flowing into the nonmagnetic metal (N) wire is absorbed into the half-metallic ferromagnetic (HMF) wire. The absorbed spin current is then converted into charge current owing to its spin-orbit interaction. 
The electric field $E_{\rm ISHE}$ is proportional to the absorbed spin current $j_s$ via the inverse spin Hall effect~\cite{Saitoh06, Valenzuela06, Kimura07},
%%%
\begin{eqnarray}
  E_{\rm ISHE} 
  &=& 
  \sigma^{-1}_{\rm SH}j_s.
\label{Eq:ISHE}
\end{eqnarray}
%%%
Here, $\sigma_{\rm SH}$ is the spin Hall conductivity. 

The absorbed spin current $j_s$ in the lateral spin valve structure corresponds to the flow of the minority-spin electrons. In the $x$ direction, the absorbed spin current $j_s$ is given by $j^x_s  = j^x_\uparrow - j^x_\downarrow$. Owing to charge conservation, the charge current $j^x_c = j^x_\uparrow + j^x_\downarrow$ in the $x$ direction vanishes. 
The absorbed spin current $j^x_s$ is then given by $j^x_s  =-2 j^x_\downarrow$. This is sufficient for the contribution from the minority-spin elections to the spin Hall conductivity to be considered relevant. 
%===
\section{Spin Hall conductivity in half-metallic ferromagnets \label{Sec:SHE}}
%===
In this section, spin Hall conductivity in a half-metallic ferromagnet is calculated by using the Kubo formula. Note that the inverse spin Hall conductivity is derived from the Onsager's reciprocal relation~\cite{Hankiewicz05}. Here, we focus on the extrinsic spin Hall effect which arises from impurity scattering. We start with the following Hamiltonian,
%%%
\begin{eqnarray}
  H &=& H_{\rm Hub} + H_{\rm imp}.
\label{Eq:Model-H}
\end{eqnarray}
%%%
Here, $H_{\rm imp}$ describes the impurity potential and is given by
%%%
\begin{eqnarray}
  H_{\rm imp} 
	&=&
	\sum_{\bmk\bmk'}\sum_{\alpha\beta}v_{\bmk-\bmk'}[\delta_{\alpha\beta}+i\eta_{\rm so}(\bmk\times\bmk')\cdot{\bm \sigma}_{\alpha\beta}] \nonumber \\
	&\times & c^{\dag}_{\bmk\alpha}c_{\bmk'\beta},
\label{Eq:Model-so}
\end{eqnarray}
%%%		
where $v_{\bmk-\bmk'}$ is the Fourier transform of the impurity potential $v_{\rm imp}\sum_{{\rm imp}\in {\rm impurities}} \delta(\bmr-\bmr_{\rm imp})$ and $\eta_{\rm so}$ measuring the strength of the spin-orbit interaction~\cite{Takahashi08, Tse06}. 

The retarded and advanced Green's functions of the minority-spin electrons without the spin-orbit scattering are written as $G^{R/A}_{\bmk\omega\downarrow}=(\omega - \varepsilon_{\bmk\downarrow}-\Sigma^{R/A}_{\bmk\downarrow})^{-1}$, where $\omega$ and $\Sigma^{R/A}_{\bmk\omega\downarrow}$ are the frequency and the self-energy of the minority-spin electrons, respectively.
%%%%%%%%%%%%%%%%%%%%%%%%%%%%
	\begin{figure}[h] 
		\begin{center}
		\scalebox{0.35}[0.35]{\includegraphics{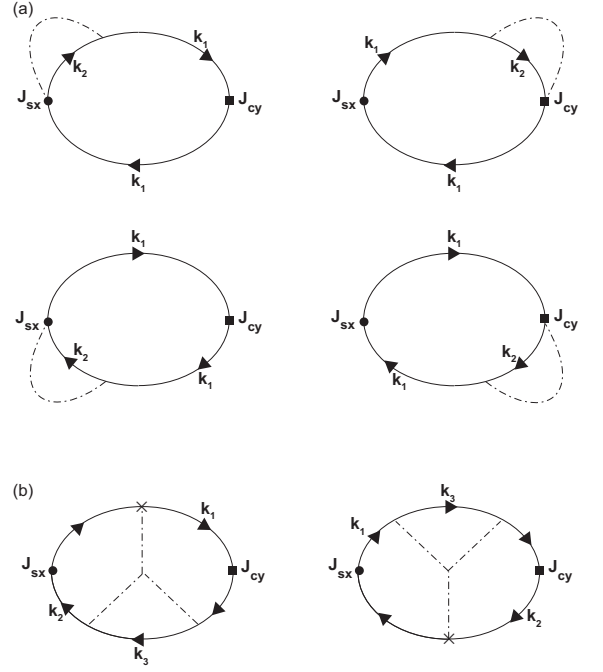}}
		\end{center}
		\caption{Diagrams for (a) the side-jump and (b) the skew-scattering contributions to spin Hall conductivity\cite{Tse06}. The solid and dashed-dotted lines represent the propagators of the minority-spin electrons and the impurity potential, respectively. The spin current $J_{\rm sx}$ and the charge current $J_{\rm cy}$ are denoted by a solid circle and a solid square, respectively. The spin-orbit interaction is denoted by a cross. 
		}
		\label{fig3_Ohnuma}
	\end{figure}
%%%%%%%%%%%%%%%%%%%%%%%%%%%
	
In the extrinsic spin Hall effect in metals, side-jump and skew-scattering mechanisms lead to contributions in spin Hall conductivity~\cite{Sinova16, Hoffmann13, Zhang00, Takahashi08, Engel05, Tse06}. 
The side-jump contribution of the minority-spin electrons $\sigma^{\rm SJ}_{\rm SH\downarrow}$, as described by Tse {\it et al}.~\cite{Tse06}, is shown in Fig.~\ref{fig3_Ohnuma}(a) and is given by
%%%
\begin{eqnarray}
	&&\sigma^{\rm SJ}_{\rm SH\downarrow}
	= \frac{e^{2}\eta_{\rm so}n_{\rm imp}v^{2}_{0}}{\pi m} \nonumber \\
	&\times &
\sum_{\bmk_1\bmk_2} [k^{2}_{1y} G^{R}_{\bmk_1\downarrow}G^{A}_{\bmk_1\downarrow}{\rm Im}G^{R}_{\bmk_2\downarrow}],
\label{Eq:SJ-Kubo-01}
\end{eqnarray}
%%%
and the skew-scattering contribution of the minority-spin electrons $\sigma^{\rm SS}_{\rm SH\downarrow}$ is shown in Fig.~\ref{fig3_Ohnuma}(b) and is  given by
%%%
\begin{eqnarray}
	&&\sigma^{\rm SS}_{\rm SH\downarrow}
	= \frac{e^{2}\hbar^{2}\eta_{\rm so}n_{\rm imp}v^{3}_{0}}{2\pi m} \nonumber \\
	&\times&\sum_{\bmk_1\bmk_2\bmk_3}  
	[k^{2}_{1x}k^{2}_{2y}G^{R}_{\bmk_1\downarrow}G^{A}_{\bmk_1\downarrow}
	G^{R}_{\bmk_2\downarrow}G^{A}_{\bmk_2\downarrow}
	{\rm Im}G^{R}_{\bmk_3\downarrow}],
\label{Eq:SS-Kubo-01}
\end{eqnarray}
%%%
where in both Eqs.~(\ref{Eq:SJ-Kubo-01}) and (\ref{Eq:SS-Kubo-01}) $G^{R/A}_{\bmk\downarrow}=(\varepsilon_{\rm F} - \varepsilon_{\bmk\downarrow}-\Sigma^{R/A}_{\bmk\downarrow})^{-1}$ represents the retarded and advanced Green's functions of the minority-spin electrons at the Fermi level, while $\Sigma^{R/A}_{\bmk\downarrow}$ is the self-energy of the minority-spin electrons at the Fermi level. Note that $\sigma^{\rm SJ}_{\rm SH\downarrow}$ and $\sigma^{\rm SS}_{\rm SH\downarrow}$ are proportional to the summation of ${\rm Im}G^{R}_{\bmk_3\downarrow}$ over $\bmk$ in Eqs.~(\ref{Eq:SJ-Kubo-01}) and (\ref{Eq:SS-Kubo-01}). This indicates that $\sigma^{\rm SJ}_{\rm SH\downarrow}$ and $\sigma^{\rm SS}_{\rm SH\downarrow}$ are sensitive to the density of states of the minority-spin electrons $N^{\downarrow} _{\varepsilon_{\rm F}}$ at the Fermi level. 

By substituting the relations $|G^R_{\bmk\downarrow}|^2	\approx1/\Delta^{2}$ and ${\rm Im}G^R_{\bmk\downarrow}\approx{\rm Im}\Sigma^R_{\bmk\downarrow}/\Delta^{2}$ into Eqs.~(\ref{Eq:SJ-Kubo-01}) and (\ref{Eq:SS-Kubo-01}), respectively, where the energy spectrum $\varepsilon_{\bmk}$ is approximated to the Fermi energy and the exchange band splitting $\Delta$ is much larger than the imaginary part of the self-energy ${\rm Im}\Sigma^R_{\bmk\downarrow}$, we obtain 
%%%
\begin{eqnarray}
	\sigma^{\rm SJ}_{\rm SH\downarrow}
	&=& -\frac{e^{2}\eta_{\rm so}\hbar n_{\rm imp}v^{2}_{0}}{\pi m\Delta^4}\frac{3n_{\uparrow}k^2_{\rm F}}{5} \nonumber \\
	&\times&
	\sum_{\bmk_2} 
	{\rm Im}\Sigma^{R}_{\bmk_2\downarrow},
\label{Eq:SJ-Kubo-02}
\end{eqnarray}
%%%
and
%%%
\begin{eqnarray}
	\sigma^{\rm SS}_{\rm SH\downarrow}
	&=& -\frac{e^{2}\hbar^{2}\eta_{\rm so}\hbar^3 n_{\rm imp}v^{3}_{0}}{2\pi m^2\Delta^6}
	\Big(\frac{3n_{\uparrow}k^2_{\rm F}}{5}\Big)^2 \nonumber \\
	&\times&
	\sum_{\bmk_3}
	{\rm Im}\Sigma^{R}_{\bmk_3\downarrow},
\label{Eq:SS-Kubo-02}
\end{eqnarray}	
%%%	
respectively. 
The isotropic Fermi surface when taking the summation over $\bmk$ has been assumed as $\sum_{\bmk_1}k^{2}_{1y}= (3n_{\uparrow}k^2_{\rm F})/5$. 

The summation of the self-energy $\sum_{\bmk}{\rm Im}\Sigma^{R}_{\bmk\downarrow}$ is divided into the following two parts:
%%%
\begin{eqnarray}
	\sum_{\bmk} {\rm Im}\Sigma^R_{\bmk\downarrow}
	&=& \sum_{\bmk} ({\rm Im}\Sigma^{R, U}_{\bmk\downarrow} + {\rm Im}\Sigma^{R, {\rm imp}}_{\bmk\downarrow}).
\label{Eq:SelfE-total-01}
\end{eqnarray}
%%%	
The first term on the right-hand side of Eq.~(\ref{Eq:SelfE-total-01}) describes the electron-magnon interaction shown in Sec.~\ref{Sec:RevHMF}. The summation of ${\rm Im}\Sigma^{R, U}_{\bmk\downarrow}$ over $\bmk$ is then taken as follows:
%%%
\begin{eqnarray}
	\sum_{\bmk} {\rm Im}\Sigma^{R, U}_{\bmk\downarrow}
	&=& -\frac{3\pi\Delta^2}{2}N^{\uparrow}_{\varepsilon_{\rm F}} \Big(\frac{k_{\rm B} T}{\hbar Dk^2_{\rm F}}\Big)^{\frac{3}{2}}\gamma_0.
\label{Eq:SelfE-U}
\end{eqnarray}
%%%
The second term on the right-hand side of Eq.~(\ref{Eq:SelfE-total-01}) describes the impurity scattering ${\rm Im}\Sigma^{R, {\rm imp}}_{\bmk\downarrow}$ and is given by ${\rm Im}\Sigma^{R, {\rm imp}}_{\bmk\downarrow}=2n_{\rm imp}\Delta^{-2}\sum_{\bmk'} v^2_{\bmk-\bmk'} {\rm Im}\Sigma^{R, U}_{\bmk-\bmk', \downarrow}$, where $n_{\rm imp}$ is the impurity density. 

The summation of ${\rm Im}\Sigma^{R, {\rm imp}}_{\bmk\downarrow}$ over $\bmk$ is taken as follows:
%%%
\begin{eqnarray}
	\sum_{\bmk} {\rm Im}\Sigma^{R,{\rm imp}}_{\bmk\downarrow}
	&=& -\pi\frac{\hbar\varepsilon_{\rm F}}{n_{\uparrow}\tau_0} N^{\uparrow}_{\varepsilon_{\rm F}} \Big(\frac{k_{\rm B} T}{\hbar Dk^2_{\rm F}}\Big)^{\frac{3}{2}}\gamma_0.
\label{Eq:SelfE-imp}
\end{eqnarray}
%%%
Here, $\tau_0=\hbar[2\pi n_{\rm imp}v^2_{\rm imp}N^{\uparrow}_{\varepsilon_{\rm F}}]^{-1}$ is the relaxation time of majority-spin electrons. 

Substituting Eqs.~(\ref{Eq:SelfE-U}) and (\ref{Eq:SelfE-imp}) into Eq.~(\ref{Eq:SelfE-total-01}) affords the self-energy as follows:
%%%
\begin{eqnarray}
	\sum_{\bmk} {\rm Im}\Sigma^R_{\bmk\downarrow}
	&=& -\pi \Big(\frac{3}{2}\Delta^2 + \frac{\hbar\varepsilon_{\rm F}}{n_{\uparrow}\tau_0}\Big) \nonumber \\
	&\times& N^{\uparrow}_{\varepsilon_{\rm F}} \Big(\frac{k_{\rm B} T}{\hbar Dk^2_{\rm F}}\Big)^{\frac{3}{2}}\gamma_0.
\label{Eq:SelfE-total-02}
\end{eqnarray}
%%%	
Substituting Eq.~(\ref{Eq:SelfE-total-02}) into Eq.~(\ref{Eq:SJ-Kubo-02}) provides the side-jump contribution to the spin Hall conductivity $\sigma^{\rm SJ}_{\rm SH\downarrow}$ as follows:
%%%
\begin{eqnarray}
	&&\sigma^{\rm SJ}_{\rm SH\downarrow}
	=
	\frac{3}{5\pi^2} \sigma^{\rm SJ}_{\uparrow} \frac{\hbar\varepsilon_{\rm F}}{\Delta^2\tau_0} \nonumber \\
	&\times&
	\Big(\frac{k_{\rm B} T}{\hbar Dk^2_{\rm F}}\Big)^{3/2} \Big(\frac{3}{2} + \frac{\hbar\varepsilon_{\rm F}}{\Delta^2\tau_0 n_{\uparrow}}\Big)\gamma_0.
\label{Eq:SJ-Kubo-fin}
\end{eqnarray}
%%%
Here, $\sigma^{\rm SJ}_{\uparrow}
=\hbar\sigma_{\uparrow}\tilde{\eta}_{\rm so}/(2\tau_0\varepsilon_{\rm F})$
is the side-jump contribution of the majority-spin electrons and $\sigma_{\uparrow}=n_{\uparrow}e^2\tau_0/m$, $\tilde{\eta}_{\rm so}$ is the dimensionless parameter measuring the strength of spin-orbit interaction $\eta_{\rm so}$ as $\tilde{\eta}_{\rm so}=\eta_{\rm so}/k^2_{\rm F}$. 

%%%%%%%%%%%%%%%%%%%%%%%%%%%%
	\begin{figure}[h] 
		\begin{center}
		\scalebox{0.28}[0.28]{\includegraphics{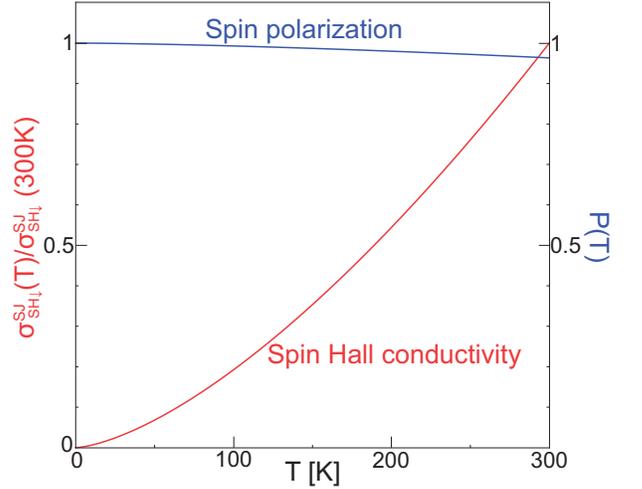}}
		\end{center}
		\caption{ (Color online)
			The temperature dependence of (a) the side-jump contribution to spin Hall conductivity $\sigma^{\rm SJ}_{\rm SH\downarrow}$ (plotted as a red line) and (b) the spin polarization $P$ (plotted as a blue line) in a half-metallic ferromagnet. Spin polarization is defined here as $P=(N^{\uparrow}_{\varepsilon_{\rm F}}-N^{\downarrow}_{\varepsilon_{\rm F}})/(N^{\uparrow}_{\varepsilon_{\rm F}}+N^{\downarrow}_{\varepsilon_{\rm F}})$~\cite{Edwards83, Irkhin06}. The data of the side-jump contribution are normalized by its value at $T=300$ K. 
		}
		\label{fig4_Ohnuma}
	\end{figure}
%%%%%%%%%%%%%%%%%%%%%%%%%%%
Figure \ref{fig4_Ohnuma} shows the temperature dependence of the side-jump contribution $\sigma^{\rm SJ}_{\rm SH\downarrow}$ and the spin polarization $P=(N^{\uparrow}_{\varepsilon_{\rm F}}-N^{\downarrow}_{\varepsilon_{\rm F}})/(N^{\uparrow}_{\varepsilon_{\rm F}}+N^{\downarrow}_{\varepsilon_{\rm F}})$~\cite{Edwards83, Irkhin06}. Here, $\sigma^{\rm SJ}_{\rm SH\downarrow}$ is proportional to $T^{3/2}$, which implies that the thermal excitation of magnons leads to the spin Hall effect of the  minority-spin electrons. It can be seen in Fig.~\ref{fig4_Ohnuma} that the spin Hall conductivity $\sigma^{\rm SJ}_{\rm SH\downarrow}$ drastically increases with temperature. On the other hand, the spin polarization $P$ is nearly constant, as the density of states of the majority-spin electrons $N^{\uparrow}_{\varepsilon_{\rm F}}$ is the dominant contribution to $P$. This indicates that an observation of the spin Hall effect may be a tool to study the electronic state of minority-spin electrons. 

Similarly, by substituting Eq.~(\ref{Eq:SelfE-total-02}) into Eq.~(\ref{Eq:SS-Kubo-02}), the skew-scattering contribution to spin Hall conductivity $\sigma^{\rm SS}_{\rm SH\downarrow}$ can be obtained as
%%%
\begin{eqnarray}
	&&\sigma^{\rm SS}_{\rm SH\downarrow}
	=
	\frac{9}{2\pi^2} \sigma^{\rm SS}_{\uparrow} \Big( \frac{\hbar\varepsilon_{\rm F}}{\Delta^2\tau_0}\Big)^2 \nonumber \\
	&\times&
	\Big(\frac{k_{\rm B} T}{\hbar Dk^2_{\rm F}}\Big)^{3/2} \Big(\frac{3}{2} + \frac{\hbar\varepsilon_{\rm F}}{\Delta^2\tau_0 n_{\uparrow}}\Big)\gamma_0,
\label{Eq:SS-Kubo-fin}
\end{eqnarray}
%%%
where $\sigma^{\rm SS}_{\uparrow}=2\pi\sigma_{\uparrow} \tilde{\eta}_{\rm so} N^{\uparrow}_{\varepsilon_{\rm F}} v_{\rm imp}/3$ is the skew-scattering contribution of the majority-spin electrons. Note that both $\sigma^{\rm SS}_{\rm SH\downarrow}$ and $\sigma^{\rm SJ}_{\rm SH\downarrow}$ are proportional to $T^{3/2}$, with $T$ being temperature in Eqs.~(\ref{Eq:SJ-Kubo-fin}) and (\ref{Eq:SS-Kubo-fin}). 
	
The ratio of the side-jump and skew-scattering contributions to spin Hall conductivity can also be demonstrated. The ratio is as follows: 
%%%
\begin{eqnarray}
	\frac{\sigma^{\rm SJ}_{\rm SH\downarrow}}{\sigma^{\rm SS}_{\rm SH\downarrow}}
	&=& 5\Big(\frac{\Delta}{\varepsilon_{\rm F}}\Big)^2 \frac{1}{v_{\rm imp} N^{\uparrow}_{\varepsilon_{\rm F}}}.
\label{Eq:SJ-SS-ratio}
\end{eqnarray}
%%%
In comparison with the ratio in normal metals~\cite{Takahashi08},
$\sigma^{\rm SJ}_{\rm SH}/\sigma^{\rm SS}_{\rm SH}
	= (3/4\pi)(\hbar/\varepsilon_{\rm F}\tau_0) (v_{\rm imp} N_{\rm F})^{-1}$, where $N_{\rm F}$ is the density of states of normal metals.  
Equation (\ref{Eq:SJ-SS-ratio}) depends on exchange band splitting $\Delta$ as $\Delta$ is much larger than the imaginary part of the self-energy ${\rm Im}\Sigma^R_{\bmk\downarrow}$, which corresponds to $\hbar/\tau_0$ in normal metals.
	
Now we discuss the dependence of spin Hall conductivity on the material parameters of a typical half-metallic ferromagnet, chromium dioxide (CrO$_2$)~\cite{Lewis97, Barry98}. First, we focus on the ratio of the spin Hall conductivity of minority-spin electrons to that of majority-spin electrons for side-jump ($\sigma^{\rm SJ}_{\rm SH\downarrow}/\sigma^{\rm SJ}_{\uparrow}$) and skew-scattering ($\sigma^{\rm SS}_{\rm SH\downarrow}/\sigma^{\rm SS}_{\uparrow}$) mechanisms. 
Since the factor $(\hbar\varepsilon_{\rm F}/\Delta^2\tau_0 n_{\uparrow})$ in Eqs.~(\ref{Eq:SJ-Kubo-fin}) and (\ref{Eq:SS-Kubo-fin}) is much smaller than $\dfrac{3}{2}$ in CrO$_2$, we have $(\sigma^{\rm SJ}_{\rm SH\downarrow}/\sigma^{\rm SJ}_{\uparrow})\approx (9/10\pi^2) 
	\gamma_0 (k_{\rm B} T/\hbar Dk^2_{\rm F})^{3/2}(\hbar\varepsilon_{\rm F}/\Delta^2 \tau_0)$ and $(\sigma^{\rm SS}_{\rm SH\downarrow}/\sigma^{\rm SS}_{\uparrow})\approx (27/4\pi^2)\gamma_0 (k_{\rm B} T/\hbar Dk^2_{\rm F})^{3/2}(\hbar\varepsilon_{\rm F}/\Delta^2\tau_0)^2$ for side-jump and skew-scattering mechanisms, respectively. 
Using the relation $\varepsilon_{\rm F} \tau_0=k_{\rm F} l/2$, with $l$ being the mean free path of half-metallic ferromagnets, we obtain $(\sigma^{\rm SJ}_{\rm SH\downarrow}/\sigma^{\rm SJ}_{\uparrow})= c_{\rm SJ}(T/T_{\rm M})^{3/2}[(\hbar/\Delta\tau_0)^2(k_{\rm F} l)]$ and $(\sigma^{\rm SS}_{\rm SH\downarrow}/\sigma^{\rm SS}_{\uparrow})= c_{\rm SS}(T/T_{\rm M})^{3/2}[(\hbar/\Delta\tau_0)^2(k_{\rm F} l)]^2$, where $c_{\rm SJ}=(9/20\pi^2)\gamma_0=0.14$, $c_{\rm SS}=(27/40\pi^2) \gamma_0=0.21$, and $T_{\rm M}\equiv \hbar Dk^2_{\rm F}/k_{\rm B}$. 
Next, we focus on the ratio of the side-jump and skew-scattering contributions to spin Hall conductivity ($\sigma^{\rm SJ}_{\rm SH\downarrow}/\sigma^{\rm SS}_{\rm SH\downarrow}$). Since $v_{\rm imp} N^{\uparrow}_{\varepsilon_{\rm F}}\approx1$ (e.g. shown in Ref.~\onlinecite{Maekawa-text}), Eq.~(\ref{Eq:SJ-SS-ratio}) is approximated as $(\sigma^{\rm SJ}_{\rm SH\downarrow}/\sigma^{\rm SS}_{\rm SH\downarrow})\approx 5(\Delta/\varepsilon_{\rm F})^2=5(\Delta\tau_0/\varepsilon_{\rm F}\tau_0)^2=5[(2\Delta\tau_0)/(\hbar k_{\rm F} l)]^2$. 
In the case of CrO$_2$~\cite{Lewis97, Barry98}, where $\Delta=1.4$ eV, $\hbar D=150$ meV\AA$^2$, $k_{\rm F}=0.93$ \AA$^{-1}$, $\tau_0=2.8\times 10^{-13}$ s, and $l=700$ \AA, we estimate $[(\hbar/\tau_0\Delta)^2(k_{\rm F} l)]=1.9\times 10^{-3}$, $T_{\rm M}=1.5\times 10^3$ K and $[(2\Delta\tau_0)/(\hbar k_{\rm F} l)]=1.8$. Then we obtain $(\sigma^{\rm SJ}_{\rm SH\downarrow}/\sigma^{\rm SJ}_{\uparrow})\approx 2.6\times 10^{-4}\times(T/1500)^{3/2}$, $(\sigma^{\rm SS}_{\rm SH\downarrow}/\sigma^{\rm SS}_{\uparrow})\approx 7.5\times 10^{-7}\times(T/1500)^{3/2}$, and $(\sigma^{\rm SJ}_{\rm SH\downarrow}/\sigma^{\rm SS}_{\rm SH\downarrow})\approx 17$.
	
%===
\section{Conclusion \label{Sec:Conclusion}}
%===
We have theoretically investigated spin transport with electron-electron correlation in a half-metallic ferromagnet. The side-jump and skew-scattering contributions to spin Hall conductivity have been derived by using the Kubo formula. Our theory explicitly manifests that the spin current is injected into a half-metallic ferromagnet at a finite temperature. The $T^{3/2}$ dependence of spin Hall conductivity for minority-spin electrons originates from the thermal excitation of magnons, i.e., the electron-electron correlation. Spin Hall conductivity is comparatively more sensitive to temperature than spin polarization. We propose that spin Hall conductivity may be a tool to study the minority-spin state in a half-metallic ferromagnet.
%%%
\acknowledgments 
%%%
We are grateful to T. Seki and Z. Wen for discussions on experimental aspects. We acknowledge Y. Sakuraba and S. Mitani for discussions on the  temperature dependence of spin polarization in half-metallic ferromagnets. This work was financially supported by the Grants-in-Aid for Scientific Research (Grant Nos. 26103006, 26247063, 16H04023, 15K05153) from JSPS and the MEXT of Japan.

%%%%%

\end{document}